\documentstyle[aps,epsfig,twocolumn,float]{revtex}
\restylefloat{figure}

\begin{document}

\title{Optimized phonon approach for 
  the diagonalization of electron-phonon problems}
\author{A.~Wei\ss e and H.~Fehske}
\address{Physikalisches Institut, Universit\"at Bayreuth, 95440 Bayreuth, Germany}
\author{G.~Wellein}
\address{Regionales Rechenzentrum Erlangen, Universit\"a{}t Erlangen, 
  91058 Erlangen, Germany}
\author{A.R.~Bishop}
\address{Theoretical Division and Center for Nonlinear Studies, 
  Los Alamos National Laboratory, Los Alamos, New Mexico 87545}
\address{~
  \parbox{14cm}{\rm 
    \medskip
    We propose a new optimized phonon approach for the
    numerical diagonalization of interacting electron-phonon systems 
    combining density-matrix and Lanczos algorithms. We demonstrate 
    the reliablity of this approach by calculating the phase diagram 
    for bi-polaron formation in the one-dimensional Holstein-Hubbard 
    model, and the Luttinger parameters for the metallic phase of the 
    half-filled one-dimensional Holstein model of spinless fermions.
}}

\maketitle

Problems of electrons or spins interacting with
lattice degrees of freedom play an important role in condensed matter
physics. To name only a few, consider for instance polaron and bi-polaron
formation in various transition metal oxides such as tungsten oxide or 
high-$T_c$ cuprates~\cite{SAL95}, Jahn-Teller effects in colossal
magnetoresistance manganites~\cite{JHSRDE97}, or Peierls and
spin-Peierls instabilities in quasi one-dimensional materials~\cite{BIJB83}. 

As a generic model for such systems the Holstein-Hubbard model,
\begin{eqnarray}\label{hholhub}
  H & = & -t \sum_{i,\sigma} (c_{i,\sigma}^{\dagger} c_{i+1,\sigma}^{} 
  + \textrm{H.c.})  
  + U \sum_{i} n_{i,\downarrow} n_{i,\uparrow}\nonumber{}\\
  & & + g\omega\sum_{i,\sigma} (b_i^{\dagger} + b_i^{}) c_{i,\sigma}^{\dagger} c_{i,\sigma}^{} 
  +\ \omega\sum_{i} b_i^{\dagger} b_i^{}\,,
\end{eqnarray}
is frequently considered, where $c_{i}^{(\dagger)}$ and $b_{i}^{(\dagger)}$ 
describe fermions and bosons on a site $i$, respectively.
In many physically relevant situations the energy scales of all the
subsystems -- electrons~($t,\ U$), phonons~($\omega$) and their 
interaction~($g\omega$) -- are of the same order of magnitude, causing 
analytic methods, and especially adiabatic techniques, to fail in most
of these cases. Even for numerical methods strong interactions are a
demanding task, since they require some cut-off in the phonon Hilbert space. 
Starting with the work of White~\cite{Whi93} in 1993, during the last 
years a class of algorithms became very popular, which based on the
use of a so called density matrix for the reduction of large Hilbert
spaces to manageable dimensions. Considerable focus has been placed
on renormalization methods for one-dimensional systems in the
thermodynamic limit. However, exact diagonalization of finite clusters
also benefit substantially from these ideas, as we will demonstrate 
in the present paper.

{\it Optimized phonon approach:} First we recapitulate the connection between 
density matrices and optimized basis states. Starting with
an arbitrary normalized quantum state 
\begin{equation}
  |\psi\rangle = \sum_{r=0}^{D_r-1} \sum_{\nu=0}^{D_\nu-1} \gamma_{\nu r} |\nu\rangle|r\rangle
\end{equation}
expressed in terms of the basis $\{|\nu\rangle|r\rangle\}$ of
the direct 
product space $H=H_\nu\otimes H_r$, we wish to reduce the dimension $D_\nu$ 
of the space $H_\nu$ by introducing a new basis, 
\begin{equation}
  |\tilde\nu\rangle = \sum_{\nu=0}^{D_\nu-1} 
  \alpha_{\tilde\nu \nu}|\nu\rangle\,,
\end{equation}
with $\tilde\nu=0\ldots (D_{\tilde\nu}-1)$ and $D_{\tilde\nu}<D_\nu$.
We call $\{|\tilde\nu\rangle\}$ an optimized basis, if the projection 
of $|\psi\rangle$ on the corresponding subspace 
$\tilde H = H_{\tilde\nu}\otimes H_r \subset H$ is as 
close as possible to the original state. Therefore we minimize
$ \| |\psi\rangle - |\tilde\psi\rangle \|^2$ with respect to the 
$\alpha_{\tilde\nu \nu}$ under the condition 
$\langle \tilde\nu'|\tilde\nu\rangle = \delta_{\tilde\nu' \tilde\nu}$, where  
\begin{equation}
  |\tilde\psi\rangle = \sum_{r=0}^{D_r-1} 
  \sum_{\tilde\nu=0}^{D_{\tilde\nu}-1} \sum_{\nu,\nu'=0}^{D_\nu-1} 
  \alpha_{\tilde\nu \nu} \alpha_{\tilde\nu \nu'}^{*} \gamma_{\nu' r} 
  |\nu\rangle|r\rangle
\end{equation}
is the projected state. Since we find
\begin{eqnarray}
  \| |\psi\rangle - |\tilde\psi\rangle \|^2
  & = & 1 - \sum_{r=0}^{D_r-1} \sum_{\tilde\nu=0}^{D_{\tilde\nu}-1} 
  \sum_{\nu,\nu'=0}^{D_\nu-1}
  \alpha_{\tilde\nu \nu} \gamma_{\nu r}^{*} 
  \gamma_{\nu' r} \alpha_{\tilde\nu \nu'}^{*} \nonumber\\
  & = & 1- \textrm{Tr}(\bbox{\alpha} \bbox{\rho} \bbox{\alpha}^{\dagger})\,,
\end{eqnarray}
where $\bbox{\rho} = \sum_{r=0}^{D_r-1} \gamma_{\nu r}^{*} \gamma_{\nu' r}$ is
called the density matrix of the state $|\psi\rangle$, we observe 
immediately that the states $\{|\tilde\nu\rangle\}$ are optimal if they are
elements of the eigenspace of $\bbox{\rho}$ corresponding to its
$D_{\tilde\nu}$ largest eigenvalues $w_{\tilde\nu}$.

Following Zhang et al.~\cite{ZJW98}, we apply these features to construct 
an optimized phonon basis for the eigenstates of an interacting 
electron/spin-phonon system. Consider a system composed of $N$ sites, 
each contributing a phonon degree of freedom 
$|\nu_i\rangle,\ \nu_i=0\ldots\infty$, and some other 
(spin or electronic) states $|r_i\rangle$. Hence, the Hilbert space
of the model under consideration is spanned by the basis
$\{\bigotimes_{i = 0}^{N-1} |\nu_i\rangle|r_i\rangle\}$. Of course,
to numerically diagonalize an Hamiltonian operating on this space, we
need to restrict ourselves to a finite-dimensional subspace. To
calculate, for instance, the lowest eigenstates of the Holstein-Hubbard 
model~(\ref{hholhub}), we could limit the phonon space spanned by
$|\nu_i\rangle = (\nu_i!)^{-1/2} (b_{i}^{\dagger})^{\nu_i}|0\rangle$ by
allowing only the states $\nu_i<D_i$. Most simply we can choose
$D_i = M\ \forall\ i$ yielding $D_{\rm ph}=M^N$ for the dimension of the
total phonon space. However, if we think of the states 
$\{\bigotimes_{i = 0}^{N-1} |\nu_i\rangle\}$ as eigenstates of the 
Hamiltonian $H_{\rm ph}=\omega\sum_{i=0}^{N-1} b_i^{\dagger} b_i$, it 
is more suitable for most problems to choose an energy cut-off instead.
Thus we used the condition $\sum_{i=0}^{N-1} \nu_i < M$, leading
to $D_{\rm ph} = {N+M-1\choose N}$, for most of our previous numerical 
work (see e.g. Ref.~\cite{BWF98}).
For weakly interacting systems already a small number $M$ of phonon 
states is sufficient to reach very good convergence for ground states
and low lying excitations. However, with increasing coupling strength most
systems require a large number of the above 'bare' phonons, thus
exceeding capacities of even large supercomputers. In some cases
one can avoid these problems by choosing an appropriate unitary
transformation of the Hamiltonian, but in general it is desirable 
to find an optimized basis automatically.

The present density-matrix algorithm~\cite{ZJW98} for the construction 
of an optimal phonon basis considered the phonon subsystem as a product 
of one large and a number of small sites. Each site except the large
one uses the same optimized basis $\{|\mu_i\rangle\} = \{|\tilde\nu\rangle\}$ 
with $\tilde\nu=0\ldots (m-1)$, while the basis of the large site consists 
of the states $\{|\tilde\nu\rangle\}$ plus some bare states 
$\{|\nu\rangle\}$, $\{|\mu_0\rangle\} = 
\textrm{ON}(\{|\tilde\nu\rangle\}\cup\{|\nu\rangle\})$, where 
$\textrm{ON}(\ldots)$ denotes orthonormalization. After a first 
initialization the optimized states are improved iteratively through
the following steps
\begin{enumerate}
\item[(1)] calculating the requested eigenstate $|\psi\rangle$ of the 
  Hamiltonian $H$ in terms of the actual basis,
\item[(2)] replacing $\{|\tilde\nu\rangle\}$ with the most important (i.e.
biggest eigenvalues $w_{\tilde\nu}$) eigenstates of the density matrix 
$\bbox{\rho}$, calculated with respect to $|\psi\rangle$ and 
$\{|\mu_0\rangle\}$,
\item[(3)] changing the additional states $\{|\nu\rangle\}$ in the set 
$\{|\mu_0\rangle\}$,
\item[(4)] orthonormalizing the set $\{|\mu_0\rangle\}$, and returning to step (1).
\end{enumerate}
A simple way to proceed in step (3) is to sweep the bare states 
$\{|\nu\rangle\}$ through a sufficiently large part of the infinite 
dimensional phonon Hilbert space. One can think of the algorithm as 
'feeding' the optimized states with bare phonons, thus allowing the 
optimized states to become increasingly perfect linear combinations of 
bare phonon states. Of course the whole procedure converges only for 
eigenstates of $H$ at the lower edge of the spectrum, since usually 
the spectrum of a Hamiltonian involving phonons has no upper bound. 
The applicability of the algorithm was demonstrated in Ref.~\cite{ZJW98} with 
the Holstein model (i.e. $U=0$ in Eq.~(\ref{hholhub})) as an example.

When we implemented the above algorithm together with a Lanczos
exact diagonalization method for our systems of interest, we found two 
objections against the above choice of an optimized basis: (i) the basis 
is not symmetric under the symmetry operations of the Hamiltonian (e.g. 
translations), and (ii) the phonon Hilbert space  is still large
($D_{\rm ph} = M\,m^{N-1}$, where $M$ is the dimension 
at the large site), since we usually need more than one 
optimized state per site. 

\begin{figure}[!htb]
  \begin{center}
    \epsfig{file= 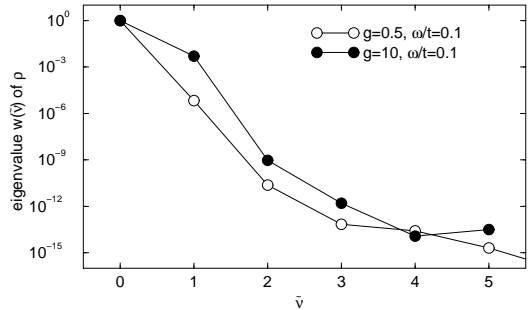, width=0.8\linewidth}
  \end{center}
  \caption{Eigenvalues $w_{\tilde\nu}$ of $\bbox{\rho}$ calculated with 
    the ground state of the Holstein model of spinless fermions for weak 
    and strong coupling. For larger $\tilde\nu$ the eigenvalues are close 
    to the numerical precision ($\approx 10^{-14}$), explaining the 
    flattening.}\label{figevals}
\end{figure}

The first problem is solved by including all those states into the 
phonon basis that can be created by symmetry operations, and by calculating
the density matrix in a symmetric way, i.e. by adding the density matrices
generated with respect to every site, not just site $i=0$.
Concerning the second problem we note that the eigenvalues $w_{\tilde\nu}$ 
of the density matrix $\bbox{\rho}$ decrease approximately exponentially, 
see Figure~\ref{figevals}. If we interpret 
$w_{\tilde\nu}\sim \xi^{\tilde\nu}$ as the probability of the system to 
occupy the corresponding optimized state $|\tilde\nu\rangle$, we immediately 
find that the probability for the complete phonon basis state 
$\bigotimes_{i = 0}^{N-1} |\tilde\nu_i\rangle$ is proportional to
$\xi^{\sum_{i = 0}^{N-1} \tilde\nu_i}$. This is reminiscent of the
energy cut-off discussed above, and we therefore propose the following
choice of phonon basis states at each site,
\begin{eqnarray}
  \forall\ i:\ \{|\mu_i\rangle\} & = & \textrm{ON}(\{|\mu\rangle\})\\ 
  |\mu\rangle & = & \left\{
    {\textrm{opt. state }|\tilde\nu\rangle,\ 0\le \mu < m\atop
      \textrm{bare state }|\nu\rangle,\ m\le \mu < M}
  \right.\,,
\end{eqnarray}
and for the complete phonon basis 
$\left\{\bigotimes_{\Sigma_i \mu_i < M} |\mu_i\rangle\right\}$,
yielding $D_{\rm ph} = {N+M-1\choose N}$.
Implementation of this optimization procedure together with our existing 
Lanczos diagonalization code~\cite{BWF98} allows the study 
of interacting electron/spin-phonon systems in a much larger parameter 
space without reaching the limits of available supercomputers. 

To demonstrate the power of the method, in the following we adress 
two frequently discussed problems: bi-polaron formation in the 
Holstein-Hubbard model, and Luttinger liquid characteristics of a 1D 
polaronic metal.

\noindent{\it (a) Bi-polarons in the Holstein-Hubbard model} 
have been the subject of numerous studies over the last decades, stimulated 
for instance by the discovery of high-$T_c$ cuprates, 
and the belief that the interplay between strong electron-phonon and 
electron-electron interactions plays a significant role in these highly 
correlated materials~\cite{AM94}. Nevertheless the influence of the Hubbard 
interaction~$U$ on bi-polaron formation is still not completely understood. 
Beside bi-polaron formation itself, an interesting open question is the 
transition between two bi-polaronic regimes, namely the inter-site and 
the on-site bi-polaron. Since the Hubbard interaction $U$ and the 
electron-phonon interaction compete, we usually need to consider 
intermediate to strong electron-phonon coupling $g$, 
or $\lambda:=g^2\omega/(2t)$, making the problem a good testing ground
for our optimized phonon algorithm.

\begin{figure}[!htb]
  \epsfig{file= 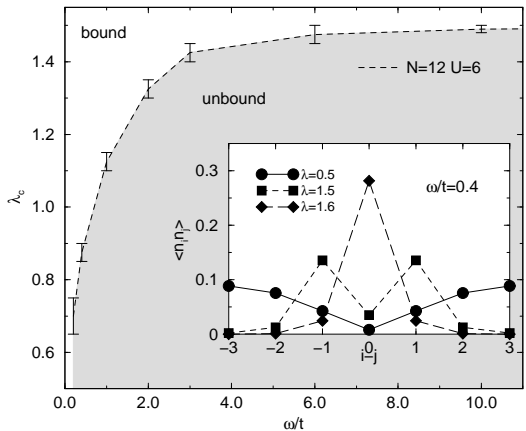, width=0.8\linewidth}
  \caption{Critical coupling for inter-site bi-polaron formation
    at fixed $U/t=6.0$ and system size $N=12$; inset: 
    $\langle n_i n_j\rangle$ correlation in the unbound,
    inter-site and on-site cases for $\omega/t=0.4$, $N=6$.}\label{figbipoPD}
\end{figure}

In a recent work Bon{\v c}a et~al.~\cite{BKT00} studied mobile bi-polarons
in the Hubbard model with the aid of a variational technique. Their focus 
is mainly on the $U$ dependence of the transition from unbound polarons to
inter-site bi-polarons and from inter-site to on-site bi-polarons at
intermediate frequencies. These transitions also show a significant 
$\omega$ dependence. Here the adiabatic frequency range is of 
special interest, since there are no appropriate analytic methods for 
small but finite frequencies. 

\begin{figure}[tb]
  \epsfig{file= 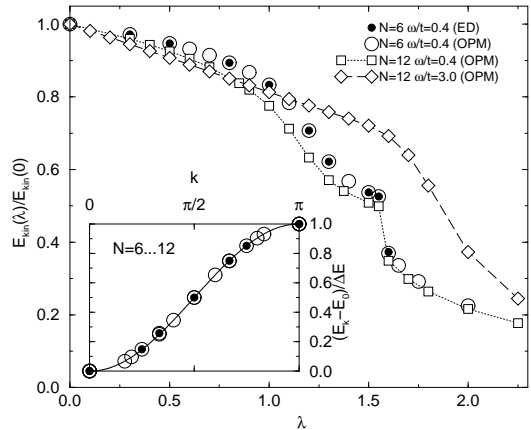, width=0.8\linewidth}
  \caption{Bi-polaron kinetic energy as a function of coupling strength
    $\lambda$ for frequencies $\omega/t=0.4$ and $3.0$, comparing optimized 
    (OPM) and bare (ED) phonons. Inset: bi-polaron dispersion at 
    $\omega/t=0.4$, $\lambda=U/(4t)$, where $\Delta E/t=0.0103$ is the
    bandwidth.
    }\label{figbipoekin}
\end{figure}

Using the optimized phonon approach on lattices sizes up to $N=12$, we 
calculated the phase diagram for the transition from unbound polarons
to inter-site bi-polarons at fixed $U$. The critical coupling 
$\lambda_c$ was determined by the condition $\Delta=0$, where $\Delta$
is the energy difference between the two-particle ground state and
twice the one-particle ground state, i.e. $\Delta=E_b - 2 E_p$. As 
indicated in Figure~\ref{figbipoPD}, the critical interaction 
$\lambda_c$ increases with frequency, reaching $U/(4t)$ as the
limiting value. The density-density correlation (see inset) signals 
a second transition from inter-site to on-site bi-polarons, which
also causes a distinct feature in the kinetic energy~\cite{FRWM95},
shown in Figure~\ref{figbipoekin} for different frequencies and system
sizes.
For $\omega=0.4t$ there is a sharp drop in $E_{\rm kin}$ at about 
$\lambda=1.6$. Near this critical coupling we observe another striking 
effect if we study the bi-polaron band dispersion: Namely, it is almost a 
perfect cosine at the critical coupling, but deviates from this
simple tight-binding dispersion for other couplings. That means in the
vicinty of $\lambda_c$ the residual bi-polaron-phonon interaction 
vanishes. At present we have no clear explanation for this free-particle 
like behaviour.
As an example we plot the rescaled dispersion for different system sizes 
and diagonalization methods in the inset of Figure~\ref{figbipoekin}. 

\noindent{\it (b) Luttinger liquid behaviour.} 
The Holstein model of spinless fermions is defined by
omitting the electron spin $\sigma$ and consequently the Coulomb 
interaction $U$ in Hamiltonian~(\ref{hholhub}). In one dimension and at half 
filling, depending on the coupling strength $g$, this model undergoes 
a transition from a gapless metallic phase to a Peierls distorted phase 
with a gap between the ground-state and lowest excitations. 
Details of this transition and the properties
of the different phases were studied with several methods over the
last years (see Refs.~\cite{HF83,ZFA89,MHM96,BMH98,WF98}). One
interesting aspect is the description of the metallic phase in terms
of an effective Luttinger model, which, according to the `Luttinger liquid
hypothesis' of Haldane~\cite{Hal80}, should be an universal picture
for the low temperature properties of all one-dimensional metals.
The two parameters of the Luttinger model, the renormalized Fermi
velocity $u_\rho$ and effective coupling constant $K_\rho$, can be
determined through the scaling behaviour of the ground-state energy
$E_0$ and the energy of charge excitations $E_{\pm 1}$ with respect 
to the system size $N$:
\begin{equation}
  {E_0(N)\over N} = \epsilon_{\infty} - {\pi u_{\rho}\over 6 N^2},\quad
  E_{\pm 1}(N) - E_0(N) = {\pi u_{\rho} \over 2 K_{\rho} N}.
\end{equation}

In a recent work~\cite{WF98} we used a variational method to 
calculate eigenstates and the resulting Luttinger parameters for
the Holstein model at half filling. Unfortunately the method
failed to give consistent results especially for $K_\rho$ in the 
anti-adiabatic regime of large frequencies $\omega\gg t$ where the Holstein
model can be well described by second order perturbation theory,
leading to an effective XXZ spin model~\cite{HF83} with known Luttinger
parameters. For large frequencies the XXZ model, as well as Monte
Carlo~\cite{MHM96} and DMRG~\cite{BMH98} calculations for the Holstein
model, yield $K_\rho<1$ corresponding to a repulsive interaction. If 
$K_\rho$, starting with the value $1$ for the noninteracting case, 
reaches ${1\over 2}$ with increasing coupling strength, the model 
undergoes a Kosterlitz-Thouless transition to a gapped phase. 
In contrast, with our variational technique we found $K_\rho>1$,
corresponding to an attractive interaction, for all frequencies.
Since it was not possible to calculate the required eigenstates 
with sufficient precision for the various system sizes needed for
finite size scaling, this unsatisfying situation could not be resolved
with our `traditional' Lanczos diagonalization procedure. 

\begin{figure}[!htb]
  \epsfig{file= 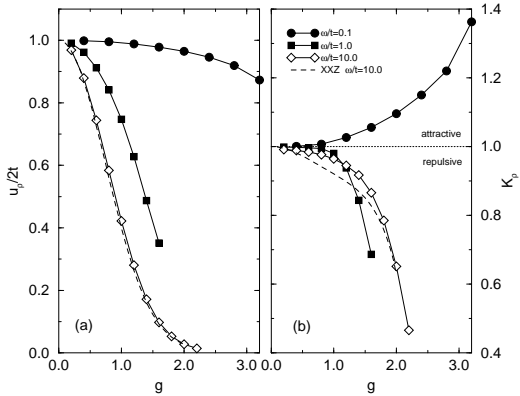, width=0.8\linewidth}
  \caption{Luttinger liquid parameters for the Holstein model of
    spinless fermions at phonon frequencies $\omega=0.1,\ 1.0$ and
    $10$.}\label{figlutt}
\end{figure}

We therefore reconsidered the problem with a more sophisticated variant
of the above phonon optimization algorithm, using two different sets 
of optimized phonon states, one for each possible fermion occupation number 
(cf. Ref.~\cite{ZJW98}). 
Together with the cut-off, this results in a further reduction 
of the Hilbert space, which is required for the diagonalization of larger 
systems. It is worth noting that this advantage is gained at the
expense of a much more complicated fermionic Hamiltonian, since every 
hopping is connected with the projection of the actual phonon state 
onto the other basis set. Hence, for other models with a more
difficult structure, like Jahn-Teller problems with two or three
phonon modes per site, this procedure is not recommended. 

In Figure~\ref{figlutt} we show the Luttinger parameters we found
by scaling the energies for system sizes up to $N=10$. 
In the anti-adiabatic frequency range the renormalized Fermi velocity $u_\rho$
is drastically supressed within the metallic phase, while for 
low phonon frequencies it remains almost unchanged up to the phase
transition. A very interesting result is the changing character of the 
interaction below $\omega\sim t$. For small frequencies the effective 
fermion-fermion interaction is attractive, while it is repulsive for 
large frequencies. Possibly there is a transition point, depending
on $g$ and $\omega$, where the model is free in lowest order. 
Further analytical studies of this behaviour will be 
reported elsewhere.

In conclusion, we have proposed an advanced phonon optimization algorithm 
for application in Lanczos diagonalization, and demonstrated its reliability 
for two strongly interacting electron-phonon systems. 

We acknowledge valuable discussion with J.~Bon{\v c}a, H.~B\"uttner, 
E.~Jeckelmann, F.~G\"ohmann, J.~Loos, and S.A.~Trugman as well as the 
provision of computer resources by NIC J\"ulich, HLR Stuttgart and 
LRZ M\"unchen. Work at Los Alamos is performed under the auspices
of the US DOE.

\end{document}